\documentclass[prd,aps,preprint,floats,nofootinbib,tightenlines,showpacs]{revtex4}

\usepackage{epsfig,amsfonts}
\begin{document} 

\pagestyle{empty}
\preprint{
\noindent
%\begin{minipage}[t]{3in}
%\begin{flushleft}
\today \\
%\end{flushleft}
% \end{minipage}
%\hfill
%\begin{minipage}[t]{3in}
%\begin{flushright}
% LBNL-- \\
%%UCB--PTH--96/xx \\
%hep-ph/\\
%December 2014
%\end{flushright}
%\end{minipage}
}

\title{Inevitable emergence of composite gauge bosons}

\author{Mahiko Suzuki}
\affiliation{
Department of Physics and Lawrence Berkeley National Laboratory\\
University of California, Berkeley, California 94720
}

\date{\today}

\begin{abstract}

A simple theorem is proved: When a gauge-invariant local field 
theory is written in terms of matter fields alone, a composite 
gauge boson or bosons must be formed dynamically. The theorem 
results from the fact that the Noether current vanishes in such 
theories. The proof is carried out by use of the charge-field 
algebra at equal time in the Heisenberg picture together with 
the well-established analyticity of the form factor of the 
current. While there is no need of diagram calculation for the 
proof, we demonstrate in the leading 1/N expansion of the 
existing models what the theorem means in diagrams and how 
the composite gauge boson emerges.  
\end{abstract}


\maketitle

\pagestyle{plain}

\setcounter{footnote}{0}

\section{Introduction}

   Some theories possess a local gauge symmetry, yet do not contain 
a gauge field explicitly. The $CP^N$ model \cite{CPN} is one of the 
examples.  It was shown in the leading 1/N expansion of the $CP^{N-1}$ 
model that a U(1) gauge boson is indeed generated as a composite 
state of matter particles.\cite{HHR} The U(1) gauge symmetry of the 
$CP^N$ model was extended by Akhmedov\cite{Akh} to the SU(2) symmetry. 
More recently, models were built with fermion matter alone.\cite{MS}
Whether the symmetry is Abelian or non-Abelian, the models with 
fermion matter cannot be reproduced by extension of the $CP^N$ 
model nor by means of the auxiliary field trick.\cite{Zel,Sak}  
Nonetheless, it was explicitly shown by the large $N$ expansion of 
the diagram calculation that these models indeed generate the 
composite gauge bosons as the massless bound states of the matter 
particles.

There is one peculiar feature common to the Lagrangian of composite 
gauge bosons.  That is, the Noether current does not exist. This can 
be shown generally as a direct consequence of local gauge invariance 
without referring to specific binding forces.\cite{MS}  In fact, 
in the case of the non-Abelian gauge theory, if the Noether current 
existed, formation of composite gauge bosons would contradict with 
the theorem of Weinberg and Witten.\cite{WW}

The diagrammatic study of the composite gauge bosons has been 
limited to the leading order of the $1/N$ expansion which amounts 
to summing up an infinite series of loop diagrams of the matter 
particles\cite{HHR,MS}.  Because of the complexity of perturbative 
computation, we cannot keep such calculation under control beyond 
the leading order of 1/N. Nonetheless, it is natural to speculate 
that the composite gauge bosons are always formed irrespectively 
of specific details of the binding force when the total Lagrangian 
is gauge invariant with matter particles alone.
 
In this paper, we attempt to prove the formation of composite gauge 
bosons to all orders of binding interactions without recourse to 
diagrams. The proof is based on the equal-time algebra of charges and 
fields in the Heisenberg picture, which incorporates all orders of 
interactions. We show that a composite gauge boson must appear 
as a pole in the form factor of the current carrying its quantum 
numbers.  Although a diagrammatic verification is redundant for 
the proof, it is reassuring and also visually helpful to 
understand the proof in terms of diagrams.  After completing our 
proof, therefore, we demonstrate in the leading 1/N expansion of 
an existing model how the statement of our theorem is realized in 
diagrams.

We organize the paper as follows: First the theorem is stated in 
Sec. II.  After the necessary input of field theory is carefully 
reviewed in Sec. III, the theorem is proved in Sec. IV with the 
equal-time algebra of charges and fields for the non-Abelian gauge 
theories of the boson matter. In Sec. V, we demonstrate in diagrams 
how the statement of the theorem is realized in the leading 1/N 
order of a concrete non-Abelian model. It is shown in Sec. VI
that the theorem holds just as well for the U(1) gauge theories.
In order to apply our argument to the fermion matter, we discuss 
in Sec. VII on an issue in the canonical quantization of the 
Dirac field, specifically, a problem related to quantization of 
constrained systems and a possibility of justifying the charge-field 
algebra without relying on the canonical quantization. We conclude 
with some perspectives in theory and phenomenology in Sec. VIII.

\section{Theorem}

The theorem is stated as follows:\\

{\em If a gauge-invariant Lagrangian field theory is written in terms 
of matter fields alone, there must be a composite gauge boson or bosons 
made of the matter particles.}\\

The gist of the theorem is that formation of the composite gauge 
boson(s) is not a possibility but the necessity. The input crucial
to prove this theorem is the absence of the Noether current in this 
class of theories.  We study the form factor of the current in the 
equal-time commutation relation of charges and fields by starting 
away from the gauge symmetry limit. Then we approach the gauge 
symmetry by continuously varying a certain parameter and prove the 
theorem without referring to diagrams or details of binding forces.    

The theorem holds in the flat space-time of (3+1) dimensions for both 
the Abelian and non-Abelian theories with boson or fermion matters. 
It is not dual to the Weinberg-Witten theorem\cite{WW}, which states 
that the non-Abelian massless gauge bosons cannot exist if the 
corresponding Lorentz-covariant conserved currents exist. Their 
theorem is mute as to whether the non-Abelian gauge bosons must 
exist or not when such currents are absent. 

\section{Non-Abelian symmetry with boson matter}

All that we use for the proof is the basic quantum field theory 
and its simple applications. To emphasize specific subtleties 
relevant to our proof, however, we give a brief review on 
elementary subjects, some of which may have fallen into 
oblivion by now.
 
\subsection{Gauge variation of Lagrangian}
The reason to discuss the spinless boson matter first is mainly 
the notational and technical simplicity related to the spins. 
But there is one complication in the canonical quantization 
of the Dirac field. Otherwise, no intrinsic difference exists 
between the boson matter and the fermion matter.
 
The Lagrangian is in the form of
\begin{equation}
    L_{tot} = \partial^{\mu}\Phi^{\dagger}\partial_{\mu}\Phi 
            -m^2\Phi^{\dagger}\Phi + L_{int}.   \label{Lb}
\end{equation}
A set of the scalar fields $\Phi/\Phi^{\dagger}$ transform locally 
like an $n/\overline{n}$-dimensional representation of a Lie group; 
\begin{equation}
 \Phi \rightarrow U \Phi,\;\; 
    \Phi^{\dagger}\rightarrow \Phi^{\dagger}U^{\dagger},  \label{gt}
\end{equation}
where $U$ is given in terms of the $n\times n$ generator matrices 
$T_a$ as
\begin{equation}
     U  =  \exp[iT_a\alpha_a(x)].  \label{U}
\end{equation}
The matrices $T_a$ obey $[T_a, T_b] = if_{abc}T_c$ with the structure
constants $f_{abc}$.

We introduce $N$ {\em copies} of the $n$-component complex scalar pairs 
$\Phi_i/\Phi_i^{\dagger}$ ($i =1,2,3 \cdots N$) since, after completing 
the proof, we make the large N-expansion in the diagram calculation 
to demonstrate how the theorem works in the explicit model.\footnote{ 
In fact, there is another reason for considering a large N. In our 
proof one-particle states will be treated as the asymptotic states. 
If confinement occurs with the composite gauge bosons, the 
one-matter-particle states are, strictly speaking, not the asymptotic 
states of the S-matrix.  The simplest way to avoid this inconvenience
is to consider the case that there exist a sufficient number of 
matter multiplets to counter the confinement.}  However, we shall
suppress the copy index $i$ hereafter unless we need to remind of it.

The interaction Lagrangian $L_{int}$ is a functional of $\Phi$,
$\Phi^{\dagger}$ and their first derivatives in the known models. 
We assume that $L_{int}$ does not contain time-derivatives of field 
higher than the first derivative. That is, $L_{int}$ should be 
just as singular as the free Lagrangian $L_0$ in regard to the 
derivatives of field.  Otherwise the gauge variation of $L_0$ 
cannot be compensated with that of $L_{int}$. \footnote{Higher 
derivatives would ruin causality in dynamics. Recall in 
classical physics that the solutions are acausal when the 
force contains a higher derivative. For instance, the radiation 
damping of a point charge. The same happens in classical field 
theory. In quantum theory we would not be able to quantize 
canonically in the Heisenberg picture if $L_{int}$ is more 
singular.} 
 
Since the free Lagrangian $L_0$ is not invariant under the local
gauge transformation Eq. (\ref{gt}), the interaction Lagrangian 
$L_{int}$ must counterbalance the gauge variation $\delta L_0$
of the free Lagrangian as 
\begin{equation}
   \delta L_{int} = - \delta L_0.   \label{deltaL}
\end{equation} 
Since $\delta L_0$ is known from the free Lagrangian in Eq. (\ref{Lb}) 
as
\begin{equation}
\delta L_0= \partial^{\mu}\Phi^{\dagger}(U^{\dagger}\partial_{\mu}U)\Phi
           +\Phi^{\dagger}(\partial^{\mu}U^{\dagger}U)\partial_{\mu}\Phi
           +\Phi^{\dagger}(\partial^{\mu}U^{\dagger}\partial_{\mu}U)\Phi,
                         \label{deltaL0}
\end{equation}
the relation of Eq. (\ref{deltaL}) determines the gauge variation 
$\delta L_{int}$ uniquely even without knowing $L_{int}$ itself.  We 
place an emphasis on this trivial but powerful constraint of gauge 
invariance since it allows us to proceed in our proof without knowing 
an explicit form of $L_{int}$.  We would need the form of $L_{int}$ 
only when we carry out, as we shall do later, a diagrammatic 
demonstration of the theorem in the interaction picture. 

Whereas we are interested in the gauge-invariant Lagrangian of
Eq. (\ref{Lb}), we insert a parameter $\lambda$ in front of $L_{int}$ as
\begin{equation}
    L^{\lambda}_{tot} = L_0 + \lambda L_{int},     \label{Llambda}
\end{equation}
and study how physics varies as $\lambda$ approaches unity. The purpose 
of this seemingly redundant procedure is the following: Since the composite 
gauge boson carries the same quantum numbers $J^{PC}=1^{--}$ as the Noether 
current, we wish to study the gauge boson through the Noether current.  
However, if we stayed exactly in the gauge symmetry limit ($\lambda = 1$), 
we would not be able to do so since the Noether current vanishes there 
according to the general theorem. (cf Appendix A.)  In order to study 
the pole of a composite gauge boson in the form factor, therefore, we 
must approach the gauge symmetry limit with $L^{\lambda}_{tot}$ of Eq. 
(\ref{Llambda}) by continuously varying the value of parameter $\lambda$ 
to $1$. By doing so, we can study where the bound-state pole of 
$J^{PC}=1^{--}$ is located off the gauge symmetry and how it moves to 
zero turning into the massless gauge boson in the gauge limit. With 
$L^{\lambda}_{tot}$ as given in Eq. (\ref{Llambda}), we approach the 
gauge limit along one special {\em path} in the functional space of 
Lagrangian. 
\footnote{Obviously there are many different ways to approach the 
gauge limit.  For instance, one may let $\lambda\rightarrow 1$ with 
the Lagrangian $L_{tot}=L_0+L_{int}+(1-\lambda)L_{br}$ where $L_{br}$ 
is some arbitrarily chosen gauge-breaking interaction. Instead we have 
chosen here the specific form $L_{tot}^{\lambda}$ for which the Noether 
current off $\lambda = 1$ takes the simple form determined by the 
free Lagrangian $L_0$ alone.}

\subsection{Noether current}
The Noether current vanishes in the gauge-symmetric field theories 
for which the Lagrangian consists only of matter fields. This is a 
simple inevitable consequence of gauge invariance, Abelian or 
non-Abelian.  Since the Noether current due to the free Lagrangian 
cannot vanish by itself, this must happen such that the contribution 
from the interaction Lagrangian cancels that from the free Lagrangian. 
The proof is very simple, as is given in Appendix A for the non-Abelian 
boson matter. Extension to other cases is trivial. 

  In short, the gauge-symmetric Lagrangian $L_{tot}$ varies under the 
infinitesimal local phase transformation by $\alpha_a(x)$ of 
Eqs. (\ref{gt}) and (\ref{U}) as
\begin{equation}
  \delta L_{tot}= i(\partial^{\mu} J_{a\mu})\alpha_a 
      +iJ_{a\mu}\;\partial^{\mu}\alpha_a +O(\alpha^2),    \label{varL}
\end{equation}
after use of the equations of motion for $\Phi$ and $\Phi^{\dagger}$ in 
the first term. Since $\alpha_a(x)$ is an arbitrary function of $x$, we 
can treat $\alpha_a(x)$ and $\partial_{\mu}\alpha_a(x)$ as independent 
of each other.  Consequently the first term of Eq. (\ref{varL}) leads 
to the definition of the Noether current and its conservation.  
The second term simply states that the Noether current must vanish.

Both $L_0$ and $L_{int}$ contribute to $J_{a\mu}$ since both contain 
the first derivatives of $\Phi$ and $\Phi^{\dagger}$ in order to 
satisfy gauge invariance.  When we modify $L_{tot}$ into
$L_0 + \lambda L_{int}$, it is no longer gauge invariant off 
$\lambda =1$ and therefore the Noether current $J^{\lambda}_{a\mu}$  
survives. It is simply given (cf Appendix A) by
\begin{equation}
    J^{\lambda}_{a\mu} = i(1-\lambda)\Big(
    \Phi^{\dagger}T_a\stackrel{\leftrightarrow}{\partial}_{\mu}\Phi
                                      \Big). \label{N}
\end{equation}
The factor $(1-\lambda)$ in front indicates the fact that the Noether 
current vanishes in the gauge limit. The Noether current thus takes 
the form identical with that of the free field theory up to the factor
$(1-\lambda)$:
\begin{equation}
  J^{free}_{a\mu}=\lim_{\lambda\rightarrow 0}
          \Big(\frac{1}{1-\lambda}J^{\lambda}_{a\mu}\Big).  \label{freeJ}
\end{equation}
However, we make a trivial but important remainder about Eq. 
(\ref{N}).  That is, 
\begin{equation}
    J^{\lambda}_{a\mu} \neq  (1-\lambda)J^{free}_{a\mu}. \label{subtlety}
\end{equation}
The reason is that when we use Eq. (\ref{N}) the fields in right-hand 
side are in the Heisenberg picture, that is, the $\Phi/\Phi^{\dagger}$ 
fields in $J^{\lambda}_{a\mu}$ incorporate all the $\lambda$-dependence 
through the interaction, while the $\Phi/\Phi^{\dagger}$ fields in 
$J^{free}_{a\mu}$ are independent of $\lambda$ ($=0$) by definition.  
It would be clearer in this respect if we wrote the fields of the 
Heisenberg picture as $\Phi(x,\lambda)$ and $\Phi^{\dagger}(x,\lambda)$. 
The implicit $\lambda$ dependence of $\Phi$ and $\Phi^{\dagger}$ in 
the Heisenberg picture incorporates all interactions and it is 
responsible for the formation of the bound states among others.
 
\subsection{Equal-time algebra of charges and fields}

We use the equal-time algebra of the charges and fields in the 
Heisenberg picture for our proof of the theorem. With the ``canonical
momentum'' defined by $\Pi\equiv\partial L/\partial(\partial^0\Phi)$, 
the field $\Phi$ obeys the equal-time commutation relation,
\begin{equation}
   [\Phi_r({\bf x},t), \Pi_s({\bf y},t)]  = i\delta_{rs}
     \delta({\bf x}-{\bf y}). \label{canonical}
\end{equation}
The subscripts $(r,s)$ refer to components of the $n$-dimensional 
representation.  Eq. (\ref{canonical}) holds separately for each of 
N copies. $\Phi^{\dagger}$ and $\Pi^{\dagger}$ obey the same form of
commutation relation, and all other equal-time commutators among 
$\Phi,\Phi^{\dagger},\Pi$ and $\Pi^{\dagger}$ vanish. In terms of 
these canonical variables, the charge component of the Noether 
current is expressed as 
\begin{eqnarray}
  J^{\lambda}_{a0} 
   &=& i(\Phi^{\dagger} T_a\Pi^{\dagger} -\Pi T_a\Phi) \nonumber\\
   &=& i(1-\lambda)(\Phi^{\dagger}T_a\stackrel{\leftrightarrow}{\partial}_0\Phi),          
                                         \label{Noe}  
\end{eqnarray}
where the summation over the N copies is understood. Notice that the
factor $(1-\lambda)$ appears when $J_{a0}$ is written in $\Phi$,
$\Phi^{\dagger}$ and their time-derivatives. But Eq. (\ref{Noe}) 
does not mean that $\Pi$ and $\Pi^{\dagger}$ are proportional to 
$1-\lambda$. (cf Appendix B) The Noether charge is defined by
\begin{equation}
    Q_a^{\lambda} = \int d^3{\bf x} J^{\lambda}_{a0}({\bf x},t). \label{Q}
\end{equation}
It is independent of time since the Noether current is conserved.  
By use of the canonical commutation relations, one can show that 
the charges form the Lie algebra,
\begin{equation}
   [Q_a^{\lambda},Q_b^{\lambda}] = if_{abc}Q_c^{\lambda}. \label{QQ}
\end{equation}
The commutation relations of $Q_a^{\lambda}$ with the fields 
$\Phi/\Phi^{\dagger}$ form the charge-field algebra,
\begin{equation}
         [Q_a^{\lambda},\Phi_r(x)] = -(T_a)_{rs}\Phi_s(x), \label{QPhi}
\end{equation}
and the hermitian conjugates.  It should be emphasized that both 
Eqs. (\ref{QQ}) and (\ref{QPhi}) are the direct consequences of
the canonical commutation relations Eq. (\ref{canonical}) and 
therefore valid irrespectively of $L_{int}$.  The peculiarity 
of the matter gauge theories to be emphasized here is that the 
Noether charge {\em operator} $Q_a^{\lambda}$ vanishes in the 
gauge symmetry limit according to Eq. (\ref{Noe}).   

Now here comes the key point. One might notice that something does 
not look quite right about Eqs. (\ref{QQ}) and (\ref{QPhi}) at least 
superficially. Let us take the matrix elements of the both sides of
Eq. (\ref{QPhi}), for instance. When the charge $Q^{\lambda}_a$ is 
expressed with the Noether current as written in the second line 
of Eq. (\ref{Noe}), it looks as if its matrix element were always
proportional to $(1-\lambda)$. If so, when it is substituted in 
Eq. (\ref{QPhi}), the left-hand side would be infinitesimally small 
like $(1-\lambda)$ near $\lambda = 1$.  On the other hand the 
matrix element of the right-hand does not vanish at $\lambda = 1$.  
The same superficial inconsistency appears as $(1-\lambda)^2$ 
{\em vs} $(1-\lambda)$ from Eq. (\ref{QQ}) too.  How should we 
answer to this question ?

There is no computational error here.  The fact that charge operator 
$Q^{\lambda}$ is proportional to $(1-\lambda)$ is a manifestation of 
the absence of the Noether current in the gauge invariant theories
that consist only of matter fields.  Then, how can the charge-field 
commutation relation Eq. (\ref{QPhi}) hold valid near $\lambda =1$ ? 
 
We shall find that this is the place where the formation of the 
composite gauge bosons enters and solves the puzzle. By examining 
the form factor of the Noether current in the following section, 
we shall find that a composite vector bound-state is formed in 
the channel of $J_{a\mu}^{\lambda}$, and therefore that the matrix 
element of
$i(\Phi^{\dagger}T_a\stackrel{\leftrightarrow}{\partial}_{\mu}\Phi)$ at 
zero momentum transfer turns out to be proportional to $1/(1-\lambda)$ 
and compensates the factor $(1-\lambda)$ in front of the operator 
$(\Phi^{\dagger}T_a\stackrel{\leftrightarrow}{\partial}_{\mu}\Phi)$.

\subsection{Dispersion relation for form factor of Noether current}

To study consistency of the powers of $(1-\lambda)$, we need to
examine the matrix elements for the both sides of Eq. (\ref{QPhi}) 
between the vacuum $\langle 0|$ and the one-particle state 
$|{\bf p}\rangle$, in particular, the one-particle matrix element 
of $J_{a\mu}^{\lambda}$ near the zero momentum-transfer limit.

    We define the Lorentz-scalar form factor $F(t,\lambda)$ by 
separating $(1-\lambda)$ from $J_{a\mu}^{\lambda}$ as
\begin{eqnarray}
 \frac{1}{1-\lambda}\langle{\bf p}',s|J_{a\mu}^{\lambda}(0)|{\bf p},r\rangle &=&
 \langle{\bf p}',s|i(\Phi^{\dagger}T_a\stackrel{\leftrightarrow}
                       \partial_{\mu}\Phi)|{\bf p},r\rangle \nonumber \\    
    &=& \sqrt{\frac{1}{4E_{{\bf p}'}E_{\bf p}}}(p'+p)_{\mu}
             (T_a)_{sr} F(t,\lambda),  \label{FFactor}
\end{eqnarray}
where the variable $t$ is the invariant momentum transfer $t=(p'-p)^2$.
Even after the factor $(1-\lambda)$ is removed from the Noether current, 
the form factor $F(t,\lambda)$ still depends on $\lambda$. This $\lambda$ 
dependence comes from the multiple interaction of $L_{int}^{\lambda}$ of 
Eq. (\ref{Llambda}), which is implicit in the Heisenberg operator
$i(\Phi^{\dagger}T_a\stackrel{\leftrightarrow}\partial_{\mu}\Phi)$, 
as we have already pointed out. 

Analyticity of the function $F(t, \lambda)$ is well known. 
$F(t, \lambda)$ is analytic 
in the variable $t$ with the branch points on the positive real axis 
of the complex $t$-plane.  The lowest branch point $t_0$ is located at 
the invariant mass squared of the lowest two-particle threshold. If 
there is a bound state of $J^{PC}=1^{--}$ with mass $m_{bound}$, 
the function $F(t, \lambda)$ has a simple pole at $m_{bound}^2$ below 
$t_0$ and, barring a tachyon, above $t=0$ for $\lambda\neq 1$. 
(See the left-side figure in Fig. 1.)
 
\noindent
\begin{figure}[ht]
\epsfig{file=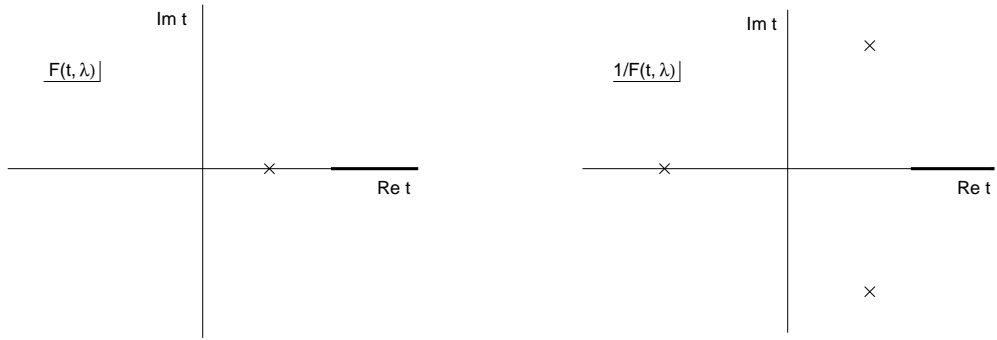,width=0.8\textwidth}
\caption{Analyticity of $F(t,\lambda)$ and $1/F(t,\lambda)$ in the complex
$t$ plane. The cross in the left-side figure indicates the pole due to 
a bound state of $J^{PC}=1^{--}$ for $F(t,\lambda)$. The crosses in the
right-side figure are due to possible poles of $1/F(t,\lambda)$, that is,
zeros of $F(t,\lambda)$.}
\label{Fig.1}
\end{figure}
 
The inverse of the form factor $1/F(t,\lambda)$ possesses the cuts 
at the same locations as $F(t,\lambda)$, but a bound-state pole of 
$F(t,\lambda)$ becomes a zero of $1/F(t,\lambda)$ and therefore does 
not generate a singularity.  The dispersion relation for 
$1/F(t,\lambda)$ therefore takes the form of \footnote{
If $F(t,\lambda)$ has a zero, it turns into a pole of $1/F(t,\lambda)$, 
which would have to be taken into account in writing the dispersion 
relation for $1/F(t,\lambda)$. Such zeros can appear in general on 
the real axis of $t$ and/or pairwise symmetrically above and below 
the real axis because of the relation $F(t,\lambda)^*=F(t^*,\lambda)$, 
where the asterisk indicates a complex conjugate.  But a zero does 
not appear for $F(t,\lambda)$ at $t=0$. The reason for $F(0,\lambda)
\neq 0$ is that $(1-\lambda)F(0,\lambda)$ is equal to the nonvanishing 
charge of the global symmetry for $\lambda \neq 1$, which must be 
nonzero.} 
\begin{equation}
  \frac{1}{F(t,\lambda)} = \frac{1}{\pi}\int_{t_0}^{\infty} 
           \frac{{\rm Im}(1/F(t',\lambda))}{t'-t-i\epsilon}dt'
   +\sum_i\frac{c_i(\lambda)}{t_i(\lambda) -t}+c_0(\lambda), \label{disp1}
\end{equation}
where $t_i(\lambda)$'s ($i=1,2,\cdots$) are the locations of zeros 
of $F(t,\lambda)$ and $c_i(\lambda)$'s are constants independent 
of $t$ with $c_0(\lambda)= 1/F(\infty,\lambda)$.  We are 
interested in the formation of a composite vector boson with small 
mass ($\rightarrow 0$ as $\lambda\rightarrow 1$), that is, a zero 
of $1/F(t,\lambda)$ on the positive real axis in the neighborhood 
of $t=0$.  Given Eq. (\ref{disp1}), we can expand $1/F(t,\lambda)$  
in the Taylor series in $t$ in the neighborhood of $t=0$ off 
$\lambda\neq 1$ as
\begin{equation}
 \frac{1}{F(t,\lambda)} = a_0(\lambda) + a_1(\lambda)t + O(t^2), 
        \;\;\; (\lambda \neq 1),             \label{expansion}
\end{equation}
where $a_0(\lambda)$ and $a_1(\lambda)$ are some real finite 
constants that may depend on $\lambda$.  Having expressed the 
behavior of $1/F(t,\lambda)$ in the form of Eq. (\ref{expansion}), 
we are ready to prove the theorem.

\section{Proof of theorem}

We take the matrix element of Eq. (\ref{QPhi}) between the vacuum 
$\langle 0|$ and the one-matter-particle state $|{\bf p},s\rangle$,
and insert a complete set of states $\sum|n\rangle\langle n|$ 
between $Q_a^{\lambda}$ and $\Phi(x)$. Since $Q_a^{\lambda}$ is a
generator of a Lie group, only the one-particle state that belongs 
to the same representation as $|{\bf p},s\rangle$ survives in 
the sum.  Use Eq. (\ref{FFactor}) to express
$\langle {\bf p},s|Q_a^{\lambda}|{\bf p},r\rangle$ in terms of the 
form factor.  We also use the relations,
\begin{eqnarray}
 \langle 0|\Phi_r(x)|{\bf p},s\rangle &=&\sqrt{\frac{1}{2E_{{\bf p}}}}
      \sqrt{Z_2}\delta_{rs}e^{-ipx}, \nonumber\\
    \langle 0|Q_a^{\lambda} &=& 0,   \label{Z2}
\end{eqnarray}
where $Z_2$ is the wave-function renormalization of the matter 
particle ($0< Z_2 < 1$). It should be emphasized that Eq. (\ref{Z2})
is valid to all orders of interaction.  After factoring out the 
group-theory coefficients and $\sqrt{Z_2}$, we are simply left with
\begin{equation}
    (1-\lambda)F(0,\lambda) = 1,
                          \label{at0}
\end{equation}
or
\begin{equation}
  F(0,\lambda) = \frac{1}{1-\lambda}.  \label{limit}
\end{equation}
This is what the charge-field algebra imposes on the form factor 
$F(t,\lambda)$ at $t=0$. Since the charge-field algebra is just
as fundamental as quantum field theory itself, the form factor 
$F(t,\lambda)$ must obey Eq. (\ref{limit}) no matter what the 
interaction of matter particles may be.

How can the form factor of $i(\Phi^{\dagger}T_a\stackrel{\leftrightarrow
}{\partial}_{\mu}\Phi)$ satisfy Eq. (\ref{limit}) ?  There must
be some dynamical reason for it.  The only possibility allowed by 
analyticity is that a bound state is present in this channel with 
the mass square proportional to $(1-\lambda)$ so that $F(t,\lambda)
\sim 1/(m_{bound}^2-t)$ near $t=0$. No other possibility exists 
according to the behavior of the form factor allowed by analyticity.

When we compare Eq.(\ref{limit}) with Eq. (\ref{expansion}), 
namely, the expansion of $1/F(t,\lambda)$ near $t=0$, we obtain
\begin{equation}
     a_0(\lambda) = 1-\lambda, \label{azero} 
\end{equation}
therefore,
\begin{equation}
  \frac{1}{F(t,\lambda)}=(1-\lambda)+a_1(\lambda)t+O(t^2).  \label{1/F} 
\end{equation}
The coefficient $a_1(\lambda)$ cannot be determined by the group 
theory alone. Eq. (\ref{1/F}) means that $F(t,\lambda)$ has 
a dynamical pole at
\begin{equation}
     t = -\frac{1-\lambda}{a_1(\lambda)}.     \label{pole}
\end{equation}
We call this pole {\em dynamical} since it is not an artifact due 
to a definition or a kinematical choice of amplitude. The value of 
$a_1(\lambda)$ that determines the location of the pole depends not 
only on $\lambda$ but also on details of the binding force. 
Therefore this pole in $t$ possesses all the properties of 
a physical bound state. It ought to be a composite vector-meson.

 Analyticity of the form factor follows from local field theory. 
With the help of analyticity, the charge-field algebra thus requires 
that a bound state be formed in the channel of $J^P=1^{--}$ with 
the mass squared proportional to $(1-\lambda)$.  When this happens, 
the multiplicative factor $1-\lambda$ of the charge operator 
$Q^{\lambda}_{a}$ coming from the Noether current is canceled by 
the dynamical factor $1/(1-\lambda)$ due to the bound-state pole 
$\sim 1/(m_{bound}^2-t)$ in $F(t,\lambda)$, where $m_{bound}^2
\propto (1-\lambda)$.  There is no other possibility. The puzzle 
is thus solved and the proof has been completed.

It should be pointed out that the crucial relation Eq. (\ref{limit})
for our proof can also be obtained in the form of 
$[(1-\lambda)F(0,\lambda)]^2=(1-\lambda)F(0,\lambda)$ by taking the 
one-particle expectation value for the both sides of the charge 
algebra Eq. (\ref{QQ}).

We add a few remarks before closing this short Section. 

The preceding argument gives us one interesting byproduct: 
Although the local Noether current vanishes in the gauge limit, 
the conserved Noether charge can still be defined for the matter 
particles through the limiting value   
$\lim_{\lambda\rightarrow 1}(1-\lambda)F(0,\lambda)$. The value 
of this charge is equal to what we would naively assign as 
the global charge to the matter particle.  It is reassuring 
that we still have the global Noether charge as the conserved 
quantum number in the gauge symmetry limit even though the Noether 
current operator itself disappears. 

Existence of the non-Abelian Noether charges as the limiting 
values has no conflict with the Weinberg-Witten theorem. To rule 
out the non-Abelian gauge-boson formation by the Weinberg-Witten 
theorem, we must have a Lorentz-covariant conserved current 
{\em density} that is capable of transferring spatial 
momentum. \cite{WW} In the gauge theories that consist only 
of matter fields, such a local current density does not exist
in the gauge symmetry limit. Therefore the global charge as 
defined above does not interfere with the Weinberg-Witten 
theorem.

Once a set of massless vector-bound states are formed in a gauge 
invariant theory, these bosons ought to be the gauge bosons of the 
underlying Lie group. The argument leading to this conclusion is, 
in short, that there is no other way known in field theory to 
accommodate such massless vector bosons in conformity with the 
gauge symmetry built in the total Lagrangian.
When the couplings of higher dimension are included, perturbative 
renormalizability does not hold in the space-time dimension of 
four. Nonetheless, when they are written in terms of effective
gauge fields, all interactions up to the dimension four are 
exactly the same as in the standard renormalizable gauge theory.  
The couplings of higher dimension for the matter fields can be 
combined and cast into gauge-invariant combinations with the 
effective vector gauge fields. The explicit demonstration was 
given through diagram computation of the higher dimensional
couplings up to the dimension six in the 1/N expansion of 
the known Abelian and non-Abelian models. \cite{MS}

\section{Diagrammatic study}

    The proof of our theorem is complete in the preceding 
section. Nothing needs to be added mathematically. Since 
the proof does not refer to any specific group property of 
the matter fields or their interactions, the theorem should 
hold for all non-Abelian gauge theories of boson matter. 
Nonetheless, it is reassuring to see that the bound-state pole 
is indeed generated in the form factor and that the pole 
migrates with the value of parameter $\lambda$ in the way 
as we have asserted. It will help us to envision the theorem 
in terms of diagrams since the diagrams often give us better 
or more intuitive understanding of physics.

For diagrammatic demonstration, we choose the SU(2) doublet 
model and make the large N expansion.  Except for keeping 
the leading 1/N terms, the diagrammatic calculation below 
makes no approximation.  To work in the large N expansion, 
we introduce the N doublets of matter. The interaction 
Lagrangian of the SU(2) gauge symmetry is given by \cite{Akh,MS} 
\begin{equation}
     L_{int}^{\lambda} = \lambda
 \frac{(\sum_i\Phi_i^{\dagger}\tau_a
 \stackrel{\leftrightarrow}{\partial}_{\mu}\Phi_i)
  (\sum_j\Phi_j^{\dagger}\tau_a\stackrel{
  \leftrightarrow}{\partial}^{\mu}\Phi_j)}{4\sum_k  \Phi^{\dagger}_k\Phi_k},               
                                              \label{L}
\end{equation}
where the summations over $i,j$ and $k$ run from $1$ to $N$. When 
the free Lagrangian of $\Phi$ and $\Phi^{\dagger}$ is added to this 
$L_{int}^{\lambda}$, the total Lagrangian $L_0 + L_{int}^{\lambda}$ 
is SU(2) gauge invariant at $\lambda = 1$.  When the value of 
$\lambda$ is in a right range, this interaction generates an SU(2) 
triplet of bound states in the channel of $J^{PC}=1^{--}$ . In the 
gauge symmetry limit, the force is just right to make the bound states 
exactly massless in the leading 1/N order.\footnote{
We should remark here that the form of $L_{int}$ appears to be unique 
up to addition of terms that are gauge invariant by themselves  
e.g., globally invariant nonderivative interactions. It is easy to 
show that such nonderivative interactions do not affect the composite 
gauge-boson mass nor coupling in the leading 1/N order.\cite{MS}} 

When we perform the diagram calculation, we express the denominator 
of Eq. (\ref{L}) in sum of its vacuum expectation value and 
normal-ordered product and expand it around the vacuum expectation 
value in the power series of the normal-ordered terms,\cite{MS} 
\begin{equation}
      L_{int}^{\lambda}  = \lambda
 \frac{(\sum_i\Phi_i^{\dagger}\tau_a\stackrel{\leftrightarrow}{\partial}_{\mu}\Phi_i)
 (\sum_j\Phi_j^{\dagger}\tau_a\stackrel{\leftrightarrow}{\partial}^{\mu}\Phi_j)}{
            4\sum_k\langle 0|\Phi_k^{\dagger}\Phi_k|0\rangle }\times
    \sum_{l=0}(-1)^l\Big(\frac{\sum_k:\!\Phi_k^{\dagger}\Phi_k\!:}{ \sum_k
    \langle 0|\Phi_k^{\dagger}\Phi_k|0\rangle}\Big)^l,  \label{normalorder}
\end{equation}
where $:\!\Phi^{\dagger}\Phi\!:$ denotes the normal-ordered product of
$\Phi^{\dagger}\Phi$. To obtain the form factor $F(t,\lambda)$ of 
$i(\Phi^{\dagger}\frac{1}{2}\tau_a\stackrel{\leftrightarrow}{\partial}_{\mu}
\Phi)$ defined in Eq. (\ref{FFactor}), we follow the leading 1/N 
computation of the two-body scattering amplitude performed in Ref. 
\cite{MS}. It amounts to iteration of the bubble diagrams, as shown 
in Fig. 2.  
\noindent
\begin{figure}[ht]
\epsfig{file=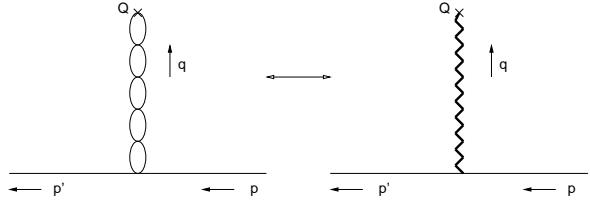,width=0.47\textwidth}
\caption{The form factor $F(t,\lambda)$ in the leading $1/N$ order
($t=q^2$). Each bubble in the left-side figure gives the function $K(t)$ 
in Eq. (\ref{iteration}) and its iteration generates a vector bound-state 
in the right-side figure.}
\label{Fig.2}
\end{figure}

After the group-theory coefficients have been factored out, the form 
factor $F(t,\lambda)$ is obtained as the solution of the simple 
algebraic equation
\begin{equation}
   F(t,\lambda) = 1 + K(t) F(t,\lambda), \label{iteration}
\end{equation}
where $K(t)$ comes from the single bubble in the left side figure of 
Fig. 2. Since we are interested in $F(t,\lambda)$ near $t=0$, we need 
$K(t)$ also near $t=0$ in Eq. (\ref{iteration}).  We carry out the 
loop integral of the bubble with the dimensional regularization to 
preserve gauge invariance.  The result is
\begin{equation}
     K(t) =\lambda\Big(1 + (1-D/2)\frac{t}{6m^2} \Big) + O(t^2),  \label{K}
\end{equation}
where $m$ is the matter-particle mass and $D$ is the space-time dimension.
With this function K(t), the inverse form factor is given by
\begin{equation}
  \frac{1}{F(t,\lambda)} = (1-\lambda)- 
  \lambda\frac{(1-D/2)t}{6m^2} +O(t^2). \label{F}
\end{equation}
This form of $1/F(t,\lambda)$ clearly shows that a vector-boson 
pole exists in $F(t,\lambda)$ and that the pole goes to zero as 
$\lambda\rightarrow1$. By comparing Eq. (\ref{F}) with the 
coefficients defined in Eq. (\ref{expansion}) in the preceding 
section, we find
\begin{eqnarray}
   a_0(\lambda) &=& 1-\lambda, \nonumber \\
   a_1(\lambda) &=& -\lambda(1-D/2)/6m^2.       \label{coeff}
\end{eqnarray}
The coefficient $a_0(\lambda)=1-\lambda$ agrees with what we have 
obtained in Eq. (\ref{azero}) in the preceding section. This is no
surprise since it is a requirement of the Noether charge being 
the generator of the global symmetry group off $\lambda = 1$.  
The coefficient $a_1(\lambda)$ determines the location of the 
bound-state pole $m_{bound}^2$ as a function of $\lambda$ and the 
matter-particle mass $m$.  As we expect, the location of the pole 
reaches zero as we approach the gauge symmetry limit, 
$\lambda\rightarrow 1$:
\begin{equation}
  m^2_{bound} = \frac{6(1-\lambda)}{\lambda(1-D/2)}m^2. \label{bmass}
\end{equation}

This exercise in the SU(2) model illustrates how our theorem works. 
While the Noether current operator disappears like $(1-\lambda)$ 
as we approach the gauge limit, the location of the bound-state pole 
converges to zero so as to cancel this $(1-\lambda)$ factor with 
$1/m_{bound}^2 \propto 1/(1-\lambda)$ at $t=0$.
 
The diagrammatic exercise presented here indicates that up to a 
proportionality constant the Noether current acts like a composite 
vector-boson field $V_{\mu}$ whose mass turns to zero in the
gauge limit.  This may remind some theorists of the field-current 
identity of Kroll, Lee, and Zumino\cite{KLZ} that identified 
the gauge current of hadrons with the (massive) gauge field. 
They attempted to equate the electromagnetic current $J_{\mu}^{EM}$ 
to the $\rho^{\circ}$-$\omega$ or $\rho^{\circ}$-$\omega$-$\phi$ 
field up to a scale factor; $J^{EM}_{\mu} = fV_{\mu}^{\rho-\omega}$.
But there is a fundamental difference. Being massive, the 
$\rho^{\circ}/\omega$ mesons are not gauge bosons of the flavor 
SU(2)$\times$U(1).  The photon being composite was not their option. 
Our passing remark here is only that if one lets $m^2_{\rho}$, 
$m^2_{\omega}\rightarrow 0$ in the field-current identity, 
such a limit has some resemblance to our matter gauge models.

Although the SU(2) matter model was shown to produce the gauge 
bosons as bound states in the leading order of 1/N expansion 
\cite{MS}, going beyond this order in the diagram calculation 
is nearly impossible because of the complexity of the nonleading 
orders. However, now that our theorem has been proved, the 
gauge-boson generation is correct to all orders of the 1/N 
expansion, that is, there is no need to do higher-order diagram 
calculation. This is one place where the power of our theorem 
should be appreciated.

We make one closing remark for this section.  Our proof turns 
out to be extremely simple primarily because the charge operator 
$Q_a^{\lambda}$ connects a one-particle state only to another 
one-particle state that belongs to the same multiplet. This
would not be the case if the momentum transfer ${\bf q}$ is 
nonvanishing across the current. The spatial Fourier components 
$Q^{\lambda}_a({\bf q},t)$ of the charge density
$J_{a0}^{\lambda}({\bf x},t)$ do not form a finite algebra: 
\begin{equation}
  [Q_a^{\lambda}({\bf q},t), \Phi({\bf q}',t)] = 
           - T_a\Phi({\bf q}+{\bf q}',t). \label{densityalgebra2}
\end{equation}
When we insert a complete set of states $\sum|n\rangle\langle n|$ 
between $Q_a^{\lambda}({\bf q},t)$ and $\Phi({\bf q}',t)$, all 
multiparticle states also contribute as long as their quantum 
numbers are right. In this case, the one-particle matrix element 
$\langle {\bf p}'|Q_a^{\lambda}({\bf q},t)|{\bf p}\rangle \sim 
1/(m^2_{bound}+|{\bf q}|^2)$ vanishes like $(1-\lambda)$ as 
$\lambda\rightarrow 1$ since ${\bf q}^2\neq 0$. Then, comparing 
the matrix elements on both sides of Eq. (\ref{densityalgebra2}), 
it may look as if our power dependence argument of $(1-\lambda)$ 
would fail like $(1-\lambda)$ {\em vs} $1$ since the 
one-particle state no longer provides $1/(1-\lambda)$ in the 
left-hand side. In this case, however, multiparticle states 
in $\sum|n\rangle\langle n|$ contribute as well without 
a constraint of energy conservation.\footnote{We end up with 
a sum rule which involves a continuum of states all the way up 
to infinite energies. Some examples using the charge density 
algebra are found in the Reference \cite{AD}. See also Reference
\cite{DG}.} In particular, the composite vector-boson enters 
the continuum and its polarization sum generates the mass 
singularity $\sim (-g_{\mu\nu}+k_{\mu}k_{\nu}/m^2_{bound})$ 
through its longitudinal polarization. \cite{W} This mass 
singularity would be canceled out if the vector-boson mass 
is generated by spontaneous symmetry breaking \cite{Russ2,MS2} 
and if the matrix elements are a set of physically observable 
scattering amplitudes. Since our matrix elements satisfy 
neither conditions, it ought to happen that the mass 
singularity proportional to $1/(1-\lambda)$ of the light 
vector composite survives and restores consistency in the 
$(1-\lambda)$ powers.  We do not attempt computation of 
the mass singularities here.

\section{U(1) gauge theories}

We can repeat our argument made for the non-Abelian theories
and show that the theorem works for the U(1) gauge theories 
as well.  Since the U(1) Noether current also vanishes in the 
gauge limit, we approach the U(1) gauge symmetry limit by 
multiplying the same parameter $\lambda$ on $L_{int}$ as we 
have done.  To avoid arbitrariness in the overall U(1) charge 
scale, we define the Noether current as
\begin{eqnarray}
   J_{\mu}^{\lambda} &=& -i\frac{\partial L^{\lambda}}{\partial^{\mu}\Phi}\Phi
  +i\Phi^{\dagger}\frac{\partial L^{\lambda}}{\partial^{\mu}\Phi^{\dagger}},
                         \nonumber \\
                    &=& i(1-\lambda) (\Phi^{\dagger}
    \stackrel{\leftrightarrow}{\partial}_{\mu}\Phi),\nonumber\\ 
   Q^{\lambda} &=& \int J_0^{\lambda}({\bf x},t)d^3{\bf x}.  \label{JU1}
\end{eqnarray}
Just as in the non-Abelian case, the factor $(1-\lambda)$ does not 
appear in $J_0^{\lambda}$ when we express it by use of $\Pi/\Pi^{\dagger}$;
\begin{equation}
   J_0^{\lambda} = i(\Phi^{\dagger}\Pi^{\dagger}-\Pi\Phi). \label{Jinpi}
\end{equation} 
Consequently the charge-field commutation relation does not have
an explicit dependence on $(1-\lambda)$;
\begin{equation}
[Q^{\lambda},\Phi({\bf x},t)] = -\Phi({\bf x},t),     \label{U1commu}
\end{equation}
in spite that $Q^{\lambda}=i(1-\lambda)\int
(\Phi^{\dagger}\stackrel{\leftrightarrow}{\partial}_0\Phi)d^3{\bf x}$. 

     We take the matrix element between the vacuum $\langle 0|$ and 
the one-particle state $|{\bf p}\rangle$ for the both sides of 
Eq. (\ref{U1commu}).  When we insert a complete set of states 
$\sum |n\rangle\langle n|$ between the $Q^{\lambda}$ and
$\Phi({\bf x},t)$, we are immediately led to 
\begin{equation}
  \langle {\bf p}|Q^{\lambda}|{\bf p}\rangle = 1. \label{fieldcommu2}
\end{equation}
The reasoning goes from here exactly as in the non-Abelian case:      
When $\langle {\bf p}'|Q^{\lambda}|{\bf p}\rangle$ is written as 
$(1-\lambda)F(t,\lambda)$ with the form factor $F(t,\lambda)$ 
of the Heisenberg operator 
$i(\Phi^{\dagger}\stackrel{\leftrightarrow}{\partial}_{\mu}\Phi)$,
Eq. (\ref{fieldcommu2}) requires that the function $F(t,\lambda)$ 
must behave like
\begin{equation}
    F(t,\lambda) \rightarrow \frac{1}{1-\lambda} + O(t) \label{U1F}
\end{equation} 
near $\lambda =1$ in the neighborhood of $t=0$.  This is realized
only if $F(t,\lambda)$ has a bound-state pole, $\mu^2/(m_{bound}^2 -t)$,
on the real axis in the complex $t$-plane and if $m_{bound}^2$ reaches 
zero at $\lambda\rightarrow 1$ as $m_{bound}^2 =\mu^2(1-\lambda)$.   

\section{Fermion matter}

The Noether theorem is based on the invariance of Lagrangian under 
the phase rotation of fields. Therefore, whether fields are 
canonically independent or not, the conserved Noether current 
consists of all the fields that enter Lagrangian,
\begin{equation}
  J_{a\mu} = -i\frac{\partial L}{\partial(\partial^{\mu}\Psi)}T_a\Psi
    +i\Psi^{\dagger}T_a\frac{\partial L}{
    \partial(\partial^{\mu}\Psi^{\dagger})}.     \label{Noef}
\end{equation}
If we want to treat $\Psi$ and $\Psi^{\dagger}$ on the equal 
footing, we may choose the free Lagrangian in the form
\begin{equation}
  L_0 = \frac{i}{2}\overline{\Psi}\stackrel{\leftrightarrow}{
       \!\not\partial}\Psi - m\overline{\Psi}\Psi,  \label{fFree}
\end{equation}
by adding a total divergence term.  With $L_{int}^{\lambda}$ added 
to this $L_0$, it may look trivial to repeat our proof for the 
boson matter to prove the theorem for the fermion matter.  But 
it is not the case.

If we formally defined the conjugate momentum by 
$\Pi=\partial L/\partial(\partial_0\Psi)$ with $L_0+L^{\lambda}_{int}$ 
and similarly for $\Pi^{\dagger}$, the Noether charge density would 
take the form of
\begin{equation}
 J_{a0}^{\lambda}=i(\Psi^{\dagger}T_a\Pi^{\dagger}-\Pi T_a\Psi),
                             \label{Jpsipi}          
\end{equation}
where $T_a=\frac{1}{2}\tau_a$ for the SU(2) doublet and 
$T_a \rightarrow 1$ for a unit U(1) charge. If we blindly imposed the 
canonical anticommutation relations by treating $(\Psi$, $\Pi$, 
$\Psi^{\dagger}, \Pi^{\dagger})$ as all independent of each other, 
it looks that we would obtain the charge-field algebra at equal 
time,
\begin{equation}
  [Q_a^{\lambda}, \Psi] = -T_a\Psi   \label{cfF}
\end{equation}
and its hermitian conjugate just as in the case of bosons. Then, 
with Eq. (\ref{cfF}), our proof for the boson models would apply 
to the fermion models with no modification. However, we encounter 
one problem: This naive derivation of Eq. (\ref{cfF}) is incorrect 
although the final result is most likely correct. There is 
a subtlety special to the canonical formalism of the Dirac 
field.\cite{Dirac0,Dirac,Schwinger2,Bergmann,HRT}. 

The problem arises from the fact that the Lagrangian of the 
Dirac field is linear in the time derivative and therefore that 
only two of those four variables above can be treated as
canonically independent. For instance, if one chooses $\Psi$ 
and $\Pi$ as independent variables, $\Psi^{\dagger}$ and 
$\Pi^{\dagger}$ are functions of $\Psi$ and $\Pi$. This turns 
the equal-time anticommutator $\{\Psi,\Psi^{\dagger}\}_+$ 
nontrivial and dependent on the interaction, in general.

In the matter gauge theories, the interaction $L_{int}$ contains 
the derivatives of field in order to counterbalance the gauge 
variation of the free Lagrangian $L_0$. In a such case, unlike
the Dirac field interacting with a nonderivative interaction, 
we do not have an option of setting $\Pi^{\dagger}=0$ by choosing 
$L_0$ asymmetric in $\Phi$ and $\Phi^{\dagger}$. Consequently 
the equal-time anticommutator between $\Psi$ and $\Psi^{\dagger}$ 
may become dependent on $L_{int}$ in general.  Although the 
prescription to determine the anticommutators has been known
when this happens, one has to go through cumbersome steps.
The canonical quantization is thus not best suited for our 
purpose in the case of the Dirac field since we would have to 
check each model one by one to make sure that the algebra Eq. 
(\ref{cfF}) is indeed valid for a given interaction.   

In some cases we can circumvent this procedure. For instance, 
in the known model of the U(1) symmetry \cite{MS}, we can remove 
the time-derivative of $\Psi^{\dagger}$ entirely and realize 
$\Pi^{\dagger}=0$ by an appropriate rewriting of the Lagrangian. 
Then the independent canonical variables are only $\Psi$ and 
$\Pi$, and they obey the simple equal-time anticommutator 
$\{\Psi,\Pi\}_+=i\delta({\bf x}-{\bf y})$.  It is interesting to 
note that in this case $\Psi$ turns out to be twice as large as 
what we would obtain formally by ignoring the interdependency of 
the variables. Since the Noether charge is given by a single term 
$J_0^{\lambda}= -i\Pi\Psi$ in the case of $\Pi^{\dagger}=0$, 
the correct charge-field algebra $[Q^{\lambda}, \Psi]=-\Psi$ 
immediately follows in the same form as that for the bosons. 
We shall describe in Appendix C how it works for the U(1) model.

In the case of the boson matter the charge-field algebra is an 
immediate consequence of the canonical quantization. In contrast, 
its derivation through the canonical quantization requires some 
knowledge of the interaction in advance in the case of the 
Dirac field. Our goal is to prove the theorem as generally 
as possible without referring to specific properties of the 
interaction or without knowing the interaction at all. For 
this purpose, it is desirable to derive the charge-field 
algebra Eq. (\ref{cfF}) in a way that does not rely on the 
canonical quantization.

In fact, a line of argument can be made to advocate validity 
of the charge-field algebra irrespectively of the interaction. 
It goes as follows: The charge-field algebra Eq. (\ref{cfF}) 
is obtained as the $O(\alpha)$ terms of the global symmetry 
rotation of the fields by angle $\alpha$,  
\begin{equation}
 e^{-iQ\alpha}\Psi(x)e^{iQ\alpha}= e^{i\alpha}\Psi(x) \label{U1finite} 
\end{equation}
for the field of a unit U(1) charge. For non-Abelian symmetries, 
$Q$ and $\alpha$ should be modified appropriately by attaching 
relevant group-component indices. Then going from Eq. (\ref{U1finite})
backward, ask what kind of operator the $Q$ can be.  The 
operator $Q$ must be a space-time independent Lorentz-scalar 
since the symmetry at $\lambda \neq 1$ is global but unbroken.
The operator $Q$ is dimensionless and has a negative charge 
parity since it generates a phase of the opposite sign for 
$\Psi^{\dagger}$ as $\Psi^{\dagger}e^{-i\alpha}$. The only possible 
candidate for $Q$ is a charge of some conserved vector current 
$J_{\mu}$.  Up to an overall proportionality constant, therefore, 
this current ought to be the Noether current that arises from the 
phase rotation of the fields. It is the only candidate that 
we have at hand. The Noether current has the right scale of 
proportionality constant since its scale is fixed by Eq. 
(\ref{U1finite}) that corresponds to the rotation per a unit 
angle of $\alpha$. 
This argument is a little wordy, but it is almost equally as 
good as the derivation based on the canonical quantization.  
It works for the boson matter too. 

Once Eq. (\ref{cfF}) has been accepted in one way or another, 
we can repeat what we have done for the boson matter.  Define 
the electric and magnetic form factors in the standard way as
\begin{eqnarray}
 \frac{1}{1-\lambda}\langle {\bf p}'|J_{a\mu}^{\lambda}(0)|{\bf p}\rangle &=&
  \langle{\bf p}'|\overline{\Psi}T_a\gamma_{\mu}\Psi|{\bf p}\rangle 
                                              \nonumber \\
 &=& \sqrt{\frac{m^2}{E_{{\bf p}'}E_{{\bf p}}}} 
   \overline{u}_{p'}T_a\bigg(\gamma_{\mu}F_1(t,\lambda) +
   \frac{i\sigma_{\mu\nu}q^{\nu}}{2m}F_2(t,\lambda)\bigg)u_{p}, \label{FFf}
\end{eqnarray}
where we have suppressed the indices for spins, copies and multiplet 
components of the fermion. Compare the one-particle matrix elements 
for the both sides of the charge algebra Eq. (\ref{cfF}) near 
$\lambda =1$. The consistency in the power of $(1-\lambda)$ on 
the both sides requires that the electric form factor $F_1(t,\lambda)$ 
must obey
\begin{equation} 
             F_1(0,\lambda) = \frac{1}{1-\lambda}.  \label{F1}
\end{equation}
It means existence of a pole of the composite gauge boson in 
$F_1(t,\lambda)$ at $t=m_{bound}^2\propto (1-\lambda)$. 
The magnetic form factor $F_2(t,\lambda)$ does not enter 
the ($q_{\mu}=0$) limit because of the kinematical factor 
$i\sigma_{\mu\nu}q^{\nu}$.  Refer to Reference \cite{MS} more for 
the Pauli term $F_2(t,\lambda)$, the dimension-five interaction, 
in the leading 1/N order.

Our proof ought to hold for any SU(2) multiplet other than the 
doublet and for any group higher than SU(2) as well, {\em if
such a model is built}. 

The diagrammatic demonstration is a little less simple for the 
fermion matter since two channels $^3S_1$ and $^3D_1$ couple to 
form the vector bound state.\cite{MS} But it is no more than 
a small technical complication.

\section{Summary and discussion}
   We can realize gauge invariance without introducing a fundamental 
vector gauge-field of any kind. In order to connect between the 
matter fields at separate space-time points in such theories, the 
interaction Lagrangian must be carefully concocted by including 
the derivatives of matter fields. In this paper we have proved that 
such matter interactions inevitably generate composite gauge bosons. 

The proof is based on the three properties: 

\noindent
(1) Most importantly, the Noether current vanishes in the gauge 
    symmetry limit of such theories. \\
(2) The equal-time charge-field algebra holds in the Heisenberg picture. \\ 
(3) The form factor of current obeys the well-established analyticity.
\vskip 0.2cm

  In our proof we have started with a globally invariant but not 
locally invariant theory ($\lambda\neq 1$) and then have approached 
the gauge symmetry by continuously varying the value of parameter 
$\lambda$. When we follow this path to the gauge symmetry, 
consistency of the charge-field 
algebra requires that a bound state must be present in the channel 
of $J^{PC}=1^{--}$ and turn massless in the gauge symmetry limit.  
The proof has been given step by step in detail for the non-Abelian 
gauge theories of the boson matter. The proof has been trivially 
extended to the Abelian theories. The theorem holds for the fermion 
matter as well. But we have cautioned about the issue that we encounter 
if we rely on the canonical quantization of the Dirac field. Our proof 
is valid to all orders of interactions since the theorem has been 
proved in the Heisenberg picture. 

  This theorem gives us another way to understand why the composite 
state of $J^{CP}=1^{--}$ cannot be massless if the Noether current 
exists: Because, if a massless bound state were formed in the presence  
of the nonvanishing Noether current, it would lead to the inconsistency 
$O(1/(1-\lambda)) = O(1)$ as $\lambda\rightarrow 1$ in the charge field 
algebra.  This observation applies to the Abelian theories equally well, 
while the theorem of Weinberg and Witten \cite{WW} is limited to the
non-Abelian theories.

The gauge boson formation was proved in the past only in the leading 
1/N order of the perturbative diagram calculation \cite{HHR,MS}. Now 
we have no need to attempt the higher-order perturbative calculation. 
With our theorem, the gauge boson formation is valid to all orders. 
This is certainly one significant advancement.  If someone succeeds 
in writing a matter gauge Lagrangian with a higher symmetry or with 
a multiplet of a higher representation within SU(2), our theorem 
guarantees that such a theory must have composite gauge bosons
before they are shown by diagrammatic computation. This is the 
main advancement.

Looking forward, some may ask how useful or relevant our theorem 
will be to phenomenology of particle physics. It is natural to wonder 
whether one can introduce in one way or another the idea of the 
composite gauge bosons into the standard model in the flat space-time 
of dimension four. At present, we have one obvious problem of group 
theory in doing so. That is, the non-Abelian models have been built 
only with the SU(2)-doublet matter particles.  This is sufficient for 
the minimal electroweak interaction of SU(2)$\times$U(1).  But what 
shall we do about the composite gluons ?  Is the so-far unsuccessful 
attempt to build a matter gauge-theory beyond the SU(2) doublet only 
for a technical reason or for a more fundamental reason ?  In the past 
we saw a few cases in which physics cannot be extended beyond SU(2).
One is the G-parity ($G=C\exp[iT_2\pi]$) of low-energy hadron physics. 
We know why it cannot.  Another is the instanton solution of the 
non-Abelian gauge theory \cite{Belavin}. This is because of the
winding number arising from mapping of the SU(2) solution onto the 
sphere ${\bf S}^3$ of the four-dimensional space-time.  Recall that 
the QCD instanton is no more than the SU(2) instantons embedded 
into the SU(3) parameter space. In our case unlike the instanton,
there seems to be nothing topological in our case. In the no-Abelian 
models so far invented, the special property of $\frac{1}{2}\tau_a$ 
for the SU(2)-doublet plays a crucial role. If an extension is 
possible beyond the SU(2)-doublet, it appears that we shall need 
a very different approach to model building. 
 
Once we have proved formation of composite gauge bosons, it is not 
necessary every time to go back to the original matter Lagrangian 
as far as the gauge boson interactions of dimension four are 
concerned. An obvious question is how to handle the effective 
interactions of dimension higher than four. This is the place 
where we expect to see difference between the elementary gauge 
bosons and the composite ones phenomenologically. It is too early 
to speculate on it.

\appendix
\section{Noether current}
We show that the Noether current is identically zero in gauge 
theories which consist only of matter fields.\cite{MS} Since this 
is the basis of our theorem, we reiterate it in the simplest way. 
We choose the non-Abelian gauge theory of boson matter as an example. 
Extension to fermion matter involves only minor modifications 
due to spins and anticommutativity.  

Gauge invariance of the action of the total Lagrangian $L_{tot}$ 
requires to the first order in $\alpha_a(x)$ 
\begin{eqnarray}
  \partial^{\mu}\Big(
  \frac{\partial L}{\partial(\partial^{\mu}\Phi)}T_a\Phi &-&
\Phi^{\dagger}T_a\frac{\partial L}{\partial(\partial^{\mu}\Phi^{\dagger})}
                 \Big)\alpha_a   \nonumber \\
&+& \Big(
\frac{\partial L}{\partial(\partial^{\mu}\Phi)}T_a\Phi
-\Phi^{\dagger}T_a\frac{\partial L}{\partial(\partial^{\mu}\Phi^{\dagger})}
    \Big)\partial^{\mu}\alpha_a  + 0(\alpha^2) = 0,  \label{gaugevar}
\end{eqnarray}
where the equation of motion has been used in the first term as 
usual. Since $\alpha_a$ are arbitrary functions of $x^{\mu}$, the 
terms proportional to $\alpha_a$ and $\partial_{\mu}\alpha_a$ must 
vanish separately in Eq. (\ref{gaugevar}). The terms proportional 
to $\alpha_a$ allow us to define the Noether current $J_a^{\mu}$ 
and lead us to its conservation:
\begin{eqnarray}
  J_{a\mu} &\equiv& -i\frac{\partial L}{\partial(\partial^{\mu}\Phi)}T_a\Phi
  +i\Phi^{\dagger}T_a\frac{\partial L}{\partial(\partial^{\mu}\Phi^{\dagger})},
                                      \label{Noether}  \\
  \partial^{\mu} J_{a\mu} &=& 0.    \label{conservation} 
\end{eqnarray}
Then the requirement that the terms proportional to $\partial^{\mu}\alpha_a$ 
be zero in Eq. (\ref{gaugevar}) is nothing other than the vanishing of the 
Noether current:
\begin{equation}
       J_{a\mu} = 0.     \label{N=0}
\end{equation}
When $L_{int}$ is multiplied with $\lambda$ and turned into $L_{int}^{\lambda}$, 
\begin{equation}
  L_{int} \rightarrow  \lambda L_{int} \equiv L_{int}^{\lambda},   \label{lambdaL}
\end{equation}
it breaks gauge invariance of the total Lagrangian $L^{\lambda}_{tot}\equiv
L_0+\lambda L_{int}$ so that the Noether current $J_{a\mu}$ no longer 
vanishes for $\lambda\neq 1$.  However, we do not need an explicit 
form of $L_{int}$ to obtain the Noether current for $\lambda \neq 1$ 
since the variation of $L_{int}$ is determined by that of the free 
Lagrangian $L_0$ alone through gauge invariance of $L_0+L_{int}$. To 
obtain the Noether current in this case, split the Lagrangian as
\begin{equation}
    L^{\lambda}_{tot} = (1-\lambda)L_0 +\lambda(L_0 + L_{int}).     \label{split}
\end{equation}
The second term does not contribute to the Noether current since it is
gauge invariant.  The Noether current arises only from the first term 
and takes the form of $(1-\lambda)$ times the Noether current due to $L_0$;
\begin{equation}
    J^{\lambda}_{a\mu} = i(1-\lambda)\Big(
 \Phi^{\dagger}T_a\stackrel{\leftrightarrow}{\partial}_{\mu}\Phi\Big). \label{NA}
\end{equation}

\section{Effect of interaction in equal-time algebras}
   The equal-time algebras of the charge $Q_a^{\lambda}$ are free of 
an explicit dependence on the factor $(1-\lambda)$.  It is because 
this factor does not appear in $Q_a^{\lambda}$ when it is written
in terms of $\Pi$ and $\Pi^{\dagger}$ instead of $\partial_0\Phi$ 
and $\partial_0\Phi^{\dagger}$.  The purposes of Appendix B is to 
show how the charge density acquires the factor $(1-\lambda)$ 
when we switch from $\Pi$ and $\Pi^{\dagger}$ to $\partial_0\Phi$ 
and $\partial_0\Phi^{\dagger}$, but that $\Pi$ nor $\Pi^{\dagger}$ 
vanishes individually as $\lambda \rightarrow 1$.

We go back to the canonical quantization rule of quantum mechanics 
in the Heisenberg picture, $[q_i, p_j] = i\delta_{ij}$, and make the 
correspondence $q_i(t)\rightarrow \Phi({\bf x},t)$ and $p_i(t)\rightarrow 
\Pi({\bf x},t)=\partial L_{tot}/\partial(\partial_0\Phi({\bf x},t))$.  
According to the standard quantization rule, a pair of the canonical 
``coordinate'' and ``momentum'' obeys the equal-time commutation relation,
\begin{equation}
 [\Phi({\bf x},t), \Pi({\bf y},t)] =i\delta({\bf x}-{\bf y}), \label{cano}
\end{equation}
and so forth. The unit matrices are to be understood in the right-hand
side of Eq. (\ref{cano}) with respect to the components of the group
indices, the copies and so forth. 

According to Eq. (\ref{Noether}), the charge density can be expressed 
as
\begin{equation}
    J^{\lambda}_{a0} 
      = i(\Phi^{\dagger}T_a\Pi^{\dagger} - \Pi T_a\Phi).
                                         \label{chargedensityAp}
\end{equation}
A factor of $(1-\lambda)$ does not appear in the right-hand side
of Eq. (\ref{chargedensityAp}). Consequently, the celebrated 
equal-time algebra of the charge densities results \cite{AD} as
\begin{equation}
     [J^{\lambda}_{a0}({\bf x},t), J^{\lambda}_{b0}({\bf y},t)]
       = if_{abc}J^{\lambda}_{c0}({\bf x},t)\delta({\bf x}-{\bf y})
                                                   \label{chargealgebraAp} 
\end{equation}
without $(1-\lambda)$. Similarly
\begin{equation}
     [J^{\lambda}_{a0}({\bf x},t), \Phi({\bf y},t)]
    = -T_a\Phi({\bf y},t)\delta({\bf x}-{\bf y}).  \label{fieldalgebraAp} 
\end{equation}

  When the Noether charge is written with $\partial_0\Phi$
and $\partial_0\Phi^{\dagger}$ instead of $\Pi$ and $\Pi^{\dagger}$, 
the factor of $(1-\lambda)$ appears. But this does not mean that 
$\Pi$ and $\Pi^{\dagger}$ are proportional to $(1-\lambda)$. 
It is interesting to see in the known model how the factor 
$(1-\lambda)$ appears in the charge density upon switching
from $\Pi$ and $\Pi^{\dagger}$ to $\partial_0\Phi$ and 
$\partial_0\Phi^{\dagger}$.

   Take the SU(2) doublet model \cite{MS} as an example.  
The interaction is given by
\begin{equation}
 L_{int}=\lambda\frac{(\Phi^{\dagger}\tau_a\stackrel{\leftrightarrow}{\partial}^{\mu}
  \Phi)(\Phi^{\dagger}\tau_a\stackrel{\leftrightarrow}{\partial}_{\mu}\Phi)}{
  4(\Phi^{\dagger}\Phi)}.  \label{PhiInt}
\end{equation}   
The momenta conjugate to $\Phi$ and $\Phi^{\dagger}$ are given by
\begin{eqnarray}
  \Pi &=& \frac{\partial L_{tot}^{\lambda}}{\partial(\partial_0\Phi)} \nonumber\\
      &=& \partial_0\Phi^{\dagger}+\lambda\frac{(\Phi^{\dagger}\tau_a
      \stackrel{\leftrightarrow}{\partial}_0\Phi)}{
      2(\Phi^{\dagger}\Phi)}\Phi^{\dagger}\tau_a ,    \label{Pi}
\end{eqnarray}
and its hermitian conjugate, respectively. Notice that neither $\Pi$ 
nor $\Pi^{\dagger}$ vanishes as $\lambda\rightarrow 1$.  However, taking 
the combination of $\Phi^{\dagger}\tau_a\Pi^{\dagger}-\Pi\tau_a\Phi$ and 
using $[\tau_a,\tau_b]=2\delta_{ab}$, we obtain
\begin{equation}
  i\Big(\Phi^{\dagger}\frac{\tau_a}{2}\Pi^{\dagger}-\Pi\frac{\tau_a}{2}\Phi\Big)=
   (1-\lambda)\Big(\Phi^{\dagger}\frac{\tau_a}{2}\stackrel{\leftrightarrow}{
   \partial}_0\Phi\Big).  \label{Piphi}
\end{equation} 
Dependence on the interaction enters the Noether current through 
$\Pi$ and $\Pi^{\dagger}$.  However, in the combination of 
$(\Phi^{\dagger}\frac{1}{2}\tau_a\Pi^{\dagger}-\Pi\frac{1}{2}\tau_a\Phi)$,
the contribution of the interaction turns out to be simply $\lambda$ times
$(\Phi^{\dagger}\frac{1}{2}\tau_a\stackrel{\leftrightarrow}{\partial}_0\Phi)$
with a minus sign. 

\section{Canonical quantization of Dirac field}

The complication in the canonical quantization of the Dirac field is 
due to the fact that the Lagrangian is linear in the time derivative 
and therefore the hermitian conjugate field $\Psi^{\dagger}$ is no 
longer canonically independent of ($\Psi$, $\Pi$) after $\Psi$ 
and $\Pi$ are chosen as the canonical variables.  This is an example 
of the so-called constrained dynamical 
systems.\cite{Dirac0,Dirac,Schwinger2,Bergmann,HRT,Jackiw}. 

Let us first recall the free Dirac field. When we choose the 
Lagrangian in the asymmetric form,
\begin{equation}
  L_0 = i\overline{\Psi}\not\!\partial\Psi-m\overline{\Psi}\Psi,
                \label{CfreeL0}
\end{equation}
we obtain $\Pi=\partial L_0/\partial(\partial_0\Psi) =i\Psi^{\dagger}$
and impose $\{\Psi,\Pi\}_+=i\delta({\bf x}-{\bf y})$ at equal time.
The canonical quantization is complete with this condition since 
$\Pi^{\dagger}=\partial L/\partial(\partial_0\Psi^{\dagger}) =0$. 
   
We may add a total divergence term to $L_0$ and antisymmetrize it 
with respect to $\partial_{\mu}\Psi$ and $\partial_{\mu}\Psi^{\dagger}$
as  
\begin{equation}
         L_0 = \frac{i}{2}\overline{\Psi}\stackrel{\leftrightarrow}{
     \not\partial}\Psi - m\overline{\Psi}\Psi.  \label{antisymL0}
\end{equation}
In this case we cannot proceed with the naive rule of quantization 
by treating both $\Psi$ and $\Psi^{\dagger}$ as independent coordinates.   
 
Let us consider the interacting Dirac fields.  We can sometimes 
circumvent the difficulty by modifying $L_{int}$ without changing
physics.  Consider the U(1) matter model \cite{MS} as an example.  
The interaction is given by
\begin{equation}
 L_{int}^{\lambda}= -\frac{i\lambda}{2}\frac{(\overline{\Psi}\gamma_{\mu}\Psi)(
   \overline{\Psi}\stackrel{\leftrightarrow}{\partial}^{\mu}\!\Psi)}{
   (\overline{\Psi}\Psi)},    \label{original}
\end{equation}
We add a total derivative term
\begin{equation}
     \Delta L_{int}^{\lambda} = -\frac{i\lambda}{2}\partial^{\mu}\Big(
  {(\overline{\Psi}\gamma_{\mu}\Psi)\log(\overline{\Psi}\Psi)\Big)},
             \label{CdeltaL}  
\end{equation}
to the original interaction Eq. (\ref{original}) and turn it into
\begin{equation}
  L_{int}^{\lambda}+\Delta L_{int}^{\lambda} 
      = -i\lambda\frac{(\overline{\Psi}\gamma_{\mu}\Psi)(
  \overline{\Psi}\partial^{\mu}\!\Psi)}{(\overline{\Psi}\Psi)}.    
                  \label{CLint}
\end{equation}
Here we have used $\partial^{\mu}(\overline{\Psi}\gamma_{\mu}\Psi)=0$.
The purpose of adding $\Delta L_{int}^{\lambda}$ is to remove the 
term $\partial_0\Psi^{\dagger}$ from the interaction.  Now the total 
Lagrangian reads
\begin{equation}
 L_{tot}^{\lambda}=i\overline{\Psi}\not\!\partial\Psi-m\overline{\Psi}\Psi
    -i\lambda\frac{(\overline{\Psi}\gamma_{\mu}\Psi)(
  \overline{\Psi}\partial^{\mu}\!\Psi)}{(\overline{\Psi}\Psi)}. \label{CTtot}
\end{equation}
Since $\Pi^{\dagger}=\partial L/\partial(\partial_0\Psi^{\dagger})=0$
for this Lagrangian, we can now choose $\Psi$ and $\Pi$ as canonically 
independent variables and treat $\Psi^{\dagger}$ as a trivial dependent 
variable, i.e., the constraint variable. 
The variable $\Pi$ defined by $\Pi=\partial L/\partial(\partial_0\Psi)$ 
with the Lagrangian of Eq. (\ref{CTtot}) turns out to be twice as large 
as what we would obtain for $\Pi$ by pretending ($\Psi$, $\Pi$, 
$\Psi^{\dagger}$, $\Pi^{\dagger}$) as all independent in the original 
Lagrangian. Since the simple canonical quantization relation 
\begin{equation}
 \{\Psi({\bf x},t),\Pi({\bf y},t)\}_+=i\delta({\bf x}-{\bf y})
                                 \label{U1trick}
\end{equation}
holds, we are led to the desired result, Eq. (\ref{cfF}) for $[Q,\Psi]$. 
Its hermitian conjugate correctly gives what we want for 
$[Q,\Psi^{\dagger}]$.

Alternatively we can choose $\Psi$ and $\Psi^{\dagger}$, instead of
$\Psi$ and $\Pi$, as the canonical variables for the original 
$L_{tot}^{\lambda}$. To do so, we must take account of the 
interdependency of the variables by making sure that Hamilton's 
equation of motion should hold correctly. The general prescriptions 
of this procedure have been discussed in length, but the case of 
the Lagrangian linear in the time-derivative can be presented 
in a compact mathematical form, which is found, for instance, 
in the lecture note, ``Constrained Quantization Without Tears'' 
by Jackiw \cite{Jackiw}.

\acknowledgments

The author thanks Professor K. Bardakci for helpful discussions 
on issues related to quantization of the constrained system.
This work was supported by the Director, Office of Science, Office of
High Energy and Nuclear Physics, Division of High Energy Physics,
of the U.S. Department of Energy under contract DE--AC02--05CH11231.

\end{document}